\documentclass[notitlepage,aps,prd,a4paper,11pt,amsfonts,amssymb,amsmath]{revtex4-1}
\pdfoutput=1
\usepackage[colorlinks]{hyperref}
\usepackage[ansinew]{inputenc}
\usepackage{color, MnSymbol}
\usepackage{graphicx}

\begin{document}
\title{Post-Newtonian parameters $\gamma$ and $\beta$ of scalar-tensor gravity\\ with a general potential}

\author{Manuel Hohmann}
\email{manuel.hohmann@ut.ee}
\affiliation{Institute of Physics, University of Tartu, Riia 142, 51014 Tartu, Estonia}

\author{Laur J\"arv}
\email{laur.jarv@ut.ee}
\affiliation{Institute of Physics, University of Tartu, Riia 142, 51014 Tartu, Estonia}

\author{Piret Kuusk}
\email{piret.kuusk@ut.ee}
\affiliation{Institute of Physics, University of Tartu, Riia 142, 51014 Tartu, Estonia}

\author{Erik Randla}
\email{erik.randla@ut.ee}
\affiliation{Institute of Physics, University of Tartu, Riia 142, 51014 Tartu, Estonia}

\begin{abstract}
We calculate the PPN parameters $\gamma$ and $\beta$ for scalar-tensor gravity with a
generic coupling function $\omega$ and  scalar potential $V$ in the Jordan conformal frame in the case of a
static spherically symmetric source. Since the potential generally introduces a
radial dependence to the effective gravitational constant as well as to $\gamma$ and $\beta$, we discuss the issue of
defining these PPN parameters and compare our expressions with previous calculations in simpler cases. We confront our results with current
observational constraints on the values of $\gamma$ and $\beta$ and
thus draw restrictions on the form of the functions $\omega$ and $V$ around their asymptotic background values.
\end{abstract}

\maketitle
\section{Introduction}\label{sec:introduction}
The discovery of the accelerated expansion of the universe and the
phenomenon of dark energy
has brought about a new surge of interest in alternatives to Einstein's general relativity in the recent years \cite{alternatives_reviews}.
One of the most simple and paradigmatic of
these is the Jordan-Brans-Dicke theory where the gravitational interaction is mediated by an extra scalar degree of freedom $\Psi$ in addition to the usual
tensor ones \cite{jordan-brans-dicke}. A more generic scalar-tensor gravity (STG) action is characterized by two arbitrary functions, the coupling function $\omega(\Psi)$ in the kinetic term of the scalar field and the potential $V(\Psi)$ \cite{stg_fathers, Nordtvedt1970, stg_books, stg_action, Flanagan:2004bz}.
In cosmology the STG models of dark energy
allow evolving effective barotropic index $\mathrm{w}$ and  dynamical crossing of the ``phantom divide'' \cite{stg_phantom_crossing} which remains a
curious possibility in the combined observational data
\cite{observations_of_evolving_w} and could be a sign pointing beyond the standard $\Lambda$CDM scenario based on general relativity.

However interesting its performance in cosmology, a viable gravitational
theory must also pass the tests on local scales, e.g., give a good account of the motions in our solar system.
 A natural framework for such a check is provided by the parameterized
post-Newtonian (PPN) formalism \cite{Will:1993ns, WillReview}.
Recent years have produced new
fascinating and precise measurements of the PPN parameters \cite{Fomalont:2009zg, Bertotti:2003rm, Hofmann, Lambert, Fienga:2011qh}.
The classic result for the parameters $\gamma$ and $\beta$ in STG with arbitrary $\omega(\Psi)$ but without a potential was obtained by Nordtvedt \cite{Nordtvedt1970}. More recently Olmo \cite{Olmo:2005hc} and
Perivolaropoulos \cite{Perivolaropoulos:2009ak} calculated the parameter $\gamma$ in the case of
the Jordan-Brans-Dicke theory (constant $\omega$) with a nonvanishing potential $V(\Psi)$.
The full equations up to the second post-Newtonian approximation for  generic $\omega(\Psi)$ and $V(\Psi)$ have been worked out in Ref. \cite{Xie:2007gq}, while Ref. \cite{Deng:2012rx} tackles light propagation up to the same PPN order for STG with a vanishing potential.
The PPN parameter $\gamma$ has also been found for scalar-tensor type theories where the scalar field has a non-standard kinetic term or possesses scalar-matter coupling \cite{ppn_nonstandard, Sen}.
The STG PPN issues are further discussed in the context of screening mechanisms
that mitigate the effects of the scalar field
\cite{ppn_screening}, and dynamical equivalence with $f(R)$ gravity \cite{ppn_fR}.

 This article fills the gap in the literature and presents the PPN parameters \(\gamma\) and \(\beta\) in the case of generic scalar-tensor theories with arbitrary functions \(\omega(\Psi)\) and \(V(\Psi)\). We display the action and field equations of these theories in Sec.~\ref{sec:stg} and derive their post-Newtonian approximation in Sec.~\ref{sec:ppn}. From this approximation we obtain a set of post-Newtonian equations which we solve in Sec.~\ref{sec:solution}. Finally in Sec.~\ref{sec:experiments} we compare our results with measurements in the solar system and obtain constraints on the so far arbitrary functions $\omega(\Psi)$ and $V(\Psi)$. We end with a conclusion in Sec.~\ref{sec:discussion}.


\section{Action functional and field equations}\label{sec:stg}
We study scalar-tensor theories of gravitation in which gravity is described by a dynamical scalar field \(\Psi\) in addition to the metric tensor \(g_{\mu\nu}\). We focus on theories which are defined by the action
\begin{equation} \label{eqn:action}
S = \frac{1}{2\kappa^2}\int_{V_4}d^4x\sqrt{-g}\left(\Psi R - \frac{\omega(\Psi)}{\Psi}\partial_{\rho}\Psi\partial^{\rho}\Psi - 2\kappa^2V(\Psi)\right) + S_m[g_{\mu\nu},\chi_m]
\end{equation}
in Brans-Dicke-like parameterization in the Jordan conformal frame~\cite{stg_books, stg_action, Flanagan:2004bz}. We have chosen units such that \(c = 1\), \(\hbar = 1\) and \(\kappa^2\) is related to the dimensionful Newtonian gravitational constant \(G_N\) via
\begin{equation} \label{eqn:kappa}
\kappa^2 = 8\pi G_N\,.
\end{equation}
The matter part of this action is given by \(S_m[g_{\mu\nu},\chi_m]\), where \(\chi_m\) collectively denotes all matter fields. The gravitational part of the action contains two free functions of the scalar field \(\Psi\): the coupling function \(\omega(\Psi)\) and the potential \(V(\Psi)\). A distinct scalar-tensor theory is defined by a particular choice of these functions. In the following we will restrict ourselves to the case \(2\omega + 3 > 0\) in order to avoid ghosts in the Einstein conformal frame~\cite{stg_books, stg_action}, as well as $V(\Psi)\geq 0$ and $\Psi>0$, but otherwise leave the functions \(\omega\) and \(V\) arbitrary.

The variation of the action~\eqref{eqn:action} with respect to the metric and the scalar field yields the gravitational field equations
\begin{eqnarray}
R_{\mu\nu} &=& \frac{1}{\Psi}\left[\kappa^2\left(T_{\mu\nu} - \frac{\omega + 1}{2\omega + 3}g_{\mu\nu}T\right) + \nabla_{\mu}\partial_{\nu}\Psi + \frac{\omega}{\Psi}\partial_{\mu}\Psi\partial_{\nu}\Psi - \frac{g_{\mu\nu}}{4\omega + 6}\frac{d\omega}{d\Psi}\partial_{\rho}\Psi\partial^{\rho}\Psi\right] \nonumber\\
&&+ \frac{\kappa^2}{\Psi}{g_{\mu\nu}}\frac{2\omega + 1}{2\omega + 3}V + g_{\mu\nu}\frac{\kappa^2}{2\omega + 3}\frac{dV}{d\Psi}\,, \label{eqn:metric}\\
\largesquare\Psi &=& \frac{1}{2\omega + 3}\left[\kappa^2T - \frac{d\omega}{d\Psi}\partial_{\rho}\Psi\partial^{\rho}\Psi + 2\kappa^2\left(\Psi\frac{dV}{d\Psi} - 2V\right)\right]\,,\label{eqn:scalar}
\end{eqnarray}
where \(\nabla_{\mu}\) denotes the covariant derivative, \(\largesquare \equiv g^{\mu\nu} \nabla_{\mu}\nabla_{\nu}\) is the d'Alembert operator and \(T_{\mu\nu}\) is the energy-momentum tensor obtained from the matter action \(S_m\). In the following section we will expand these field equations in a post-Newtonian approximation.


\section{Post-Newtonian approximation}\label{sec:ppn}
We now expand the field equations~\eqref{eqn:metric} and~\eqref{eqn:scalar} of scalar-tensor gravity as displayed in the preceding section up to the first post-Newtonian order. For this purpose we make use of the parameterized post-Newtonian (PPN) formalism~\cite{Will:1993ns, WillReview} and assume that the gravitating source matter is constituted by a perfect fluid which obeys the post-Newtonian hydrodynamics. In this section we start from this assumption and assign appropriate orders of magnitude to all terms appearing in the field equations. The resulting equations can then be solved subsequently for each order of magnitude in the next section.

The starting point of our calculation is the energy-momentum tensor of a perfect fluid with rest energy density \(\rho\), specific internal energy \(\Pi\), pressure \(p\) and four-velocity \(u^{\mu}\), which takes the form
\begin{equation} \label{eqn:tmunu}
T^{\mu\nu} = (\rho + \rho\Pi + p)u^{\mu}u^{\nu} + pg^{\mu\nu}\,.
\end{equation}
The four-velocity \(u^{\mu}\) is normalized by the metric \(g_{\mu\nu}\), so that \(u^{\mu}u^{\nu}g_{\mu\nu} = -1\). A basic ingredient of the PPN formalism is the perturbative expansion of all dynamical quantities in orders \(\mathcal{O}(n) \propto |\vec{v}|^n\) of the velocity \(v^{i} = u^{i}/u^0\) of the source matter in a given frame of reference. For the metric \(g_{\mu\nu}\) this is an expansion around a flat Minkowski background,
\begin{equation}\label{eqn:metricexp}
g_{\mu\nu} = \eta_{\mu\nu} + h_{\mu\nu} = \eta_{\mu\nu} + h^{(1)}_{\mu\nu} + h^{(2)}_{\mu\nu} + h^{(3)}_{\mu\nu} + h^{(4)}_{\mu\nu} + \mathcal{O}(5)\,,
\end{equation}
where each term \(h^{(n)}_{\mu\nu}\) is of order \(\mathcal{O}(n)\). In order to describe the motion of test bodies in the lowest post-Newtonian approximation an expansion up to the fourth velocity order \(\mathcal{O}(4)\) is sufficient. A detailed analysis shows that not all components of the metric need to be expanded to the fourth velocity order, while others vanish due to Newtonian energy conservation or time reversal symmetry. The only relevant, non-vanishing components of the metric perturbations are given by
\begin{equation}
h^{(2)}_{00}\,, \qquad h^{(2)}_{ij}\,, \qquad h^{(3)}_{0j}\,, \qquad h^{(4)}_{00}\,.
\end{equation}
In order to determine these components for a given matter source we must assign velocity orders also to the rest mass density, specific internal energy and pressure of the perfect fluid. Based on their orders of magnitude in the solar system one assigns velocity orders \(\mathcal{O}(2)\) to \(\rho\) and \(\Pi\) and \(\mathcal{O}(4)\) to \(p\). The energy-momentum tensor~\eqref{eqn:tmunu} can then be expanded in the form
\begin{subequations}\label{eqn:energymomentum}
\begin{eqnarray}
T_{00} &=& \rho\left(1 + \Pi + v^2 - h^{(2)}_{00}\right) + \mathcal{O}(6)\,,\\
T_{0j} &=& -\rho v_j + \mathcal{O}(5)\,,\\
T_{ij} &=& \rho v_iv_j + p\delta_{ij} + \mathcal{O}(6)\,.
\end{eqnarray}
\end{subequations}
We further assume that the gravitational field is quasi-static, so that changes are only induced by the motion of the source matter. Time derivatives \(\partial_0\) of the metric components and other fields are therefore weighted with an additional velocity order \(\mathcal{O}(1)\).

In order to apply the PPN formalism to the scalar-tensor gravity theory detailed in section~\ref{sec:stg} we further need to expand the scalar field \(\Psi\) in a post-Newtonian approximation. For this purpose we expand \(\Psi\) around its cosmological background value \(\Psi_0\),
\begin{equation}
\Psi = \Psi_0 + \psi = \Psi_0 + \psi^{(2)} + \psi^{(4)} + \mathcal{O}(6)\,,
\end{equation}
where we assume \(\Psi_0\) to be of order \(\mathcal{O}(0)\) and the perturbations \(\psi^{(n)}\) are of order \(\mathcal{O}(n)\). We also need to expand the functions \(\omega(\Psi)\) and \(V(\Psi)\) around \(\Psi_0\). A Taylor expansion to the required order takes the form
\begin{subequations}
\begin{eqnarray}
\omega &=& \omega_0 + \omega_1\psi + \mathcal{O}(\psi^2)\,,\\
V &=& V_0 + V_1\psi + V_2\psi^2 + V_3\psi^3 + \mathcal{O}(\psi^4)\,,
\end{eqnarray}
\end{subequations}
with constant expansion coefficients which we assume to be of velocity order \(\mathcal{O}(0)\). With these definitions the field equations~\eqref{eqn:metric} and~\eqref{eqn:scalar} in the lowest velocity order \(\mathcal{O}(0)\) read
\begin{eqnarray}
0 &=& \frac{\kappa^2}{2\omega_0 + 3}\left(\frac{2\omega_0 + 1}{\Psi_0}V_0 + V_1\right)\eta_{\mu\nu}\,,\\
0 &=& \frac{2\kappa^2}{2\omega_0 + 3}(\Psi_0V_1 - 2V_0)\,.
\end{eqnarray}
In order for these to be satisfied we must set \(V_0 = V_1 = 0\), which is a consequence of our expansion~\eqref{eqn:metricexp} of the metric around a flat background. The scalar field equation up to fourth velocity order then reduces to
\begin{eqnarray}
\label{eqn:scalarppn}
\left(\nabla^2 - \frac{4\kappa^2\Psi_0V_2}{2\omega_0 + 3}\right)\left(\psi^{(2)} + \psi^{(4)}\right)&=&\frac{\kappa^2}{2\omega_0 + 3}(3p - \rho - \rho\Pi) + \frac{2\kappa^2\omega_1}{(2\omega_0 + 3)^2}\rho\psi^{(2)}\nonumber\\
&&+ \psi^{(2)}_{,00} + \frac{2\kappa^2\Psi_0}{2\omega_0 + 3}\left(3V_3 - \frac{4\omega_1V_2}{2\omega_0 + 3}\right)\left(\psi^{(2)}\right)^2 - \frac{\omega_1}{2\omega_0 + 3}\psi^{(2)}_{,i}\psi^{(2)}_{,i}\nonumber\\
&&+ h^{(2)}_{ij}\psi^{(2)}_{,ij} + \left(h^{(2)}_{ij,j} + \frac{1}{2}h^{(2)}_{00,i} - \frac{1}{2}h^{(2)}_{jj,i}\right)\psi^{(2)}_{,i} + \mathcal{O}(6)\,.
\end{eqnarray}
We further expand the RHS of the gravitational tensor field equation~\eqref{eqn:metric} up to the necessary velocity order and obtain
\begin{subequations}\label{eqn:riccippn}
\begin{eqnarray}
R_{00} &=& \frac{\kappa^2}{\Psi_0(2\omega_0 + 3)}\Bigg[(\omega_0 + 2)\rho - 2\Psi_0V_2\left(\psi^{(2)} + \psi^{(4)}\right) - (\omega_0 + 2)h_{00}^{(2)}\rho \nonumber\\
&&+ \left((2\omega_0 + 3)v^2 + (\omega_0 + 2)\Pi + (3\omega_0 + 3)\frac{p}{\rho}\right)\rho + 2\Psi_0V_2h^{(2)}_{00}\psi^{(2)} \nonumber\\
&&+ \left[\left(\frac{4\Psi_0\omega_1}{2\omega_0 + 3} - 2\omega_0 - 1\right)V_2 - 3\Psi_0V_3\right]\left(\psi^{(2)}\right)^2 + \frac{\omega_1}{2\kappa^2}\psi^{(2)}_{,i}\psi^{(2)}_{,i} \nonumber\\
&&- \left(\frac{\omega_0 + 2}{\Psi_0} + \frac{\omega_1}{2\omega_0 + 3}\right)\rho\psi^{(2)}\Bigg] + \frac{1}{\Psi_0}\left(\psi^{(2)}_{,00} + \frac{1}{2}h_{00,i}^{(2)}\psi^{(2)}_{,i}\right) + \mathcal{O}(6)\,, \label{eqn:r00ppn}\\
R_{0j} &=& \frac{1}{\Psi_0}\left(\psi^{(2)}_{,0j} - \kappa^2v_j\rho\right) + \mathcal{O}(5)\,, \label{eqn:r0jppn}\\
R_{ij} &=& \frac{\kappa^2}{\Psi_0}\left[\frac{(\omega_0 + 1)\rho + 2\Psi_0V_2\psi^{(2)}}{2\omega_0 + 3}\delta_{ij} + \frac{\psi^{(2)}_{,ij}}{\kappa^2}\right] + \mathcal{O}(4)\,. \label{eqn:rijppn}
\end{eqnarray}
\end{subequations}
In order to solve these equations we finally need to fix a gauge for the metric tensor. A useful choice for the class of scalar-tensor theories we consider is given by~\cite{Nutku:1969}
\begin{subequations}
\begin{eqnarray}
h_i{}^j{}_{,j} - \frac{1}{2}h_{\mu}{}^{\mu}{}_{,i} & = & \frac{1}{\Psi_0}\psi_{,i} \,,\\
h_0{}^j{}_{,j} - \frac{1}{2}h_j{}^j{}_{,0} & = & \frac{1}{\Psi_0}\psi_{,0} \,.
\end{eqnarray}
\end{subequations}
In this gauge we can express the Ricci tensor up to the necessary velocity order as
\begin{subequations}\label{eqn:riccigauge}
\begin{eqnarray}
\label{eqn:r00gauge}
R_{00} &=& -\frac{1}{2}\nabla^2h^{(2)}_{00} - \frac{1}{2}\nabla^2h^{(4)}_{00} + \frac{1}{\Psi_0}\psi^{(2)}_{,00} \\
&&+ \frac{1}{2\Psi_0}h^{(2)}_{00,j}\psi^{(2)}_{,j} - \frac{1}{2}h^{(2)}_{00,j}h^{(2)}_{00,j} + \frac{1}{2}h^{(2)}_{jk}h^{(2)}_{00,jk} + \mathcal{O}(6)\,, \nonumber\\
R_{0j} &=& -\frac{1}{2}\nabla^2h^{(3)}_{0j} - \frac{1}{4}h^{(2)}_{00,0j} + \frac{1}{\Psi_0}\psi^{(2)}_{,0j} + \mathcal{O}(5)\,, \label{eqn:r0jgauge}\\
R_{ij} &=& -\frac{1}{2}\nabla^2h^{(2)}_{ij} +\frac{1}{\Psi_0}\psi^{(2)}_{,ij} + \mathcal{O}(4)\,. \label{eqn:rijgauge}
\end{eqnarray}
\end{subequations}
This completes our derivation of the post-Newtonian field equations. We can now use equation~\eqref{eqn:scalarppn} to solve for the scalar field perturbation \(\psi\) and equations~\eqref{eqn:riccippn} and~\eqref{eqn:riccigauge} to solve for the metric perturbations \(h_{\mu\nu}\), up to the necessary velocity orders. This will be done in the following section.


\section{Static spherically symmetric solution}\label{sec:solution}
In the previous section we have derived the field equations of scalar-tensor gravity in a post-Newtonian approximation. It is now our aim to construct a simple solution to these equations, which corresponds to the static, spherically symmetric gravitational field of a single, point-like mass. Our derivation consists of three steps. First, we solve the field equations for the metric perturbation \(h^{(2)}_{00}\) in the Newtonian approximation in Sec.~\ref{subsec:newton}. Using this result we can then solve for \(h^{(2)}_{ij}\) in Sec.~\ref{subsec:gamma} and for \(h^{(4)}_{00}\) in Sec.~\ref{subsec:beta}.


\subsection{Newtonian approximation}\label{subsec:newton}
In the remainder of this article we restrict ourselves to the gravitational field generated by a point-like mass \(M\), which is described by the energy-momentum tensor~\eqref{eqn:energymomentum} with
\begin{equation}\label{eqn:pointmass}
\rho = M\delta(\vec{x})\,, \qquad \Pi = 0\,, \qquad p = 0\,, \qquad v_i = 0\,.
\end{equation}
This simple matter source induces a static and spherically symmetric metric, which can most easily be expressed using isotropic spherical coordinates. In the rest frame of the gravitating mass we use the ansatz
\begin{subequations}\label{eqn:gmunu}
\begin{eqnarray}
g_{00} &=& -1 + 2G_{\mathrm{eff}}(r)U(r) - 2G_{\mathrm{eff}}^2(r)\beta(r)U^2(r) + \Phi^{(4)}(r) + \mathcal{O}(6)\,, \label{eqn:g00}\\
g_{0j} &=& \mathcal{O}(5)\,,\\
g_{ij} &=& \left[1 + 2G_{\mathrm{eff}}(r)\gamma(r)U(r)\right]\delta_{ij} + \mathcal{O}(4)\,, \label{eqn:gij}
\end{eqnarray}
\end{subequations}
where \(r\) denotes the radial coordinate and the Newtonian potential \(U(r)\) is given by
\begin{equation}
U(r) = \frac{\kappa^2}{8\pi}\frac{M}{r}\,.
\end{equation}
In the potential \(\Phi^{(4)}\) we collect terms of order \(\mathcal{O}(4)\) which are not of the form \(G_{\mathrm{eff}}^2\beta U^2\), such as the gravitational self-energy. The three unknown functions we need to determine are the effective gravitational constant \(G_{\mathrm{eff}}(r)\) and the PPN parameters \(\gamma(r)\) and \(\beta(r)\). The latter two can be defined either as the coefficients of the effective gravitational potential \(U_{\mathrm{eff}} = G_{\mathrm{eff}}U\) as shown in the metric~\eqref{eqn:gmunu} or as the coefficients \(\gamma_{\mathrm{eff}} = G_{\mathrm{eff}}\gamma\) and \(\beta_{\mathrm{eff}} = G_{\mathrm{eff}}^2\beta\) of the Newtonian potential terms \(U\) and \(U^2\). The first definition invokes the interpretation that the measured values of \(\gamma\) and \(\beta\) can be related to the effective gravitational potential \(U_{\mathrm{eff}}\), while the second definition suggests to relate the measured values of \(\gamma_{\mathrm{eff}}\) and \(\beta_{\mathrm{eff}}\) to the Newtonian potential \(U\) of a fixed mass \(M\). We choose the first definition in this article since the mass of the Sun, which dominates the solar system physics, is determined from its gravitational effects on the planetary motions.

For our calculation of the Newtonian limit we start by solving equation~\eqref{eqn:scalarppn} up to the second velocity order,
\begin{equation}\label{eqn:psi2}
\left(\nabla^2 - m_{\psi}^2\right)\psi^{(2)} = -\frac{\kappa^2}{2\omega_0 + 3}\rho
\end{equation}
for the scalar field perturbation \(\psi^{(2)}\). The constant
\begin{equation}\label{eqn:mpsi}
m_{\psi} = 2\kappa\sqrt{\frac{\Psi_0V_2}{2\omega_0 + 3}}
\end{equation}
can be interpreted as the mass of the scalar field \cite{Faraoni:2009km}. Equation~\eqref{eqn:psi2} is a screened Poisson equation, which is solved by
\begin{equation}
\psi^{(2)}(r) = \frac{2}{2\omega_0 + 3}U(r)e^{-m_{\psi}r}\,.
\end{equation}
The solution for \(\psi^{(2)}(r)\) thus takes the form of a Yukawa potential.

In the next step we consider equations~\eqref{eqn:r00ppn} and~\eqref{eqn:r00gauge} up to the second velocity order. From these we derive
\begin{equation}
\nabla^2h_{00}^{(2)} = \frac{m_{\psi}^2}{\Psi_0}\left(\psi^{(2)} - \frac{\omega_0 + 2}{2\Psi_0V_2}\rho\right)\,.
\end{equation}
We can write the solution in the form
\begin{equation} \label{eqn:h00}
h_{00}^{(2)}(r) = 2G_{\mathrm{eff}}(r)U(r)\,,
\end{equation}
where the effective gravitational constant is given by
\begin{equation} \label{eqn:geff}
G_{\mathrm{eff}}(r) = \frac{1}{\Psi_0}\left(1 + \frac{e^{-m_{\psi}r}}{2\omega_0 + 3}\right)\,.
\end{equation}
In order to interpret this result for \(G_{\mathrm{eff}}\) as an effective gravitational constant we need to choose an experiment in which the gravitational interaction takes place at a constant scale \(r = r_0\). We can then choose units in which \(G_{\mathrm{eff}}(r_0) = 1\). This corresponds to a rescaling of the cosmological background value \(\Psi_0\) of the scalar field to
\begin{equation}
\Psi_0 = 1 + \frac{e^{-m_{\psi}r_0}}{2\omega_0 + 3}\,.
\end{equation}
However, we cannot make this choice globally, and hence cannot remove the factor \(G_{\mathrm{eff}}(r)\) from the metric~\eqref{eqn:gmunu} by a choice of units in which \(G_{\mathrm{eff}} \equiv 1\), as it is conventionally done in the basic PPN formalism~\cite{Will:1993ns}. This is the reason for the ambiguity in the definition of the PPN parameters \(\gamma\) and \(\beta\) we discussed above.


\subsection{PPN parameter $\gamma(r)$}\label{subsec:gamma}
We now turn our focus to the spatial components \(h^{(2)}_{ij}\) of the metric perturbation. From equations~\eqref{eqn:rijppn} and~\eqref{eqn:rijgauge} we obtain
\begin{equation}
\nabla^2h^{(2)}_{ij} = -\frac{m_{\psi}^2}{\Psi_0}\left(\psi^{(2)} + \frac{\omega_0 + 1}{2\Psi_0V_2}\rho\right)\delta_{ij}\,.
\end{equation}
From this equation we can immediately read off that the solution for \(h^{(2)}_{ij}\) is diagonal. We find that
\begin{equation}\label{eqn:hij}
h_{ij}^{(2)}(r) = \frac{2}{\Psi_0}\left(1 - \frac{e^{-m_\psi r}}{2\omega_0 + 3}\right)\delta_{ij}U(r)\,.
\end{equation}
Comparison with equation~\eqref{eqn:gij} then yields the PPN parameter \(\gamma(r)\) and we obtain
\begin{equation} \label{eqn:gamma}
\gamma(r) = \frac{2\omega_0 + 3 - e^{-m_{\psi}r}}{2\omega_0 + 3 + e^{-m_{\psi}r}}\,.
\end{equation}
This result agrees with a previously obtained result for a purely quadratic scalar potential \(V(\Psi)\) and constant coupling function \(\omega(\Psi)\)~\cite{Olmo:2005hc,Perivolaropoulos:2009ak}. In the limit \(V_2 \to 0\) and fixed finite \(\omega_0\), which implies \(m_{\psi} \to 0\), the PPN parameter \(\gamma\) approaches the known value
\begin{equation}
\gamma = \frac{\omega_0 + 1}{\omega_0 + 2}
\end{equation}
for scalar-tensor gravity with a massless scalar field~\cite{Nordtvedt1970}. Analogously in the limit \(1/(2\omega_0 + 3) \to 0\) and fixed finite \(V_2\) we have \(m_{\psi} \to 0\) but find the limiting value \(\gamma = 1\). The same value \(\gamma = 1\) is also approached in the opposite limiting case of a massive scalar field with \(m_{\psi}r \gg 1\).


\subsection{PPN parameter $\beta(r)$}\label{subsec:beta}
We finally calculate the PPN parameter beta as given in equation~\eqref{eqn:g00}. For this purpose we need the fourth order perturbation \(\psi^{(4)}\) of the scalar field \(\Psi\). This can be obtained from the fourth order part of equation~\eqref{eqn:scalarppn}, which reads
\begin{eqnarray}
\left(\nabla^2 - m_{\psi}^2\right)\psi^{(4)} &=& \frac{\kappa^2}{2\omega_0 + 3}(3p - \rho\Pi) + \frac{m_\psi^2\omega_1}{2\Psi_0V_2(2\omega_0 + 3)}\rho\psi^{(2)}\nonumber\\
&&+ \psi^{(2)}_{,00} + m_\psi^2\left(\frac{3V_3}{2V_2} - \frac{2\omega_1}{2\omega_0 + 3}\right)\left(\psi^{(2)}\right)^2 - \frac{\omega_1}{2\omega_0 + 3}\psi^{(2)}_{,i}\psi^{(2)}_{,i}\\
&&+ h^{(2)}_{ij}\psi^{(2)}_{,ij} + \left(h^{(2)}_{ij,j} + \frac{1}{2}h^{(2)}_{00,i} - \frac{1}{2}h^{(2)}_{jj,i}\right)\psi^{(2)}_{,i}\,.\nonumber
\end{eqnarray}
As for the second order perturbation \(\psi^{(2)}\) we obtain a screened Poisson equation, but with a different source term. The terms involving the pressure \(p\) and specific internal energy \(\Pi\) drop out because of our choice~\eqref{eqn:pointmass} of a point mass as the source of the gravitational field. We further neglect the term \(\rho\psi^{(2)}\), which corresponds to a gravitational self-energy. Finally, the term \(\psi_{,00}\) drops out since we consider only static solutions. The solution to the remaining equation is then given by
\begin{eqnarray}
\psi^{(4)}(r) &=& \frac{2}{(2\omega_0 + 3)^2}\left(\frac{1}{\Psi_0} - \frac{\omega_1}{2\omega_0 + 3}\right)U^2(r)e^{-2m_{\psi}r} \nonumber\\
&&+ \frac{2m_{\psi}}{\Psi_0(2\omega_0 + 3)}U^2(r)r\left[e^{m_{\psi}r}\mathrm{Ei}(-2m_{\psi}r) - e^{-m_{\psi}r}\ln(m_{\psi}r)\right] \nonumber\\
&&+ \frac{3m_{\psi}}{(2\omega_0 + 3)^2}\left(\frac{V_3}{V_2} - \frac{1}{\Psi_0} - \frac{\omega_1}{2\omega_0 + 3}\right) U^2(r)r\left[e^{m_{\psi}r}\mathrm{Ei}(-3m_{\psi}r) - e^{-m_{\psi}r}\mathrm{Ei}(-m_{\psi}r)\right]\,.
\label{psi4sol}\end{eqnarray}
Here \(\mathrm{Ei}\) denotes the exponential integral, which is defined by
\begin{equation}
\mathrm{Ei}(-x) = -\int_{x}^{\infty} \frac{e^{-t}}{t} dt\,.
\end{equation}
Using the result (\ref{psi4sol}) we can now determine \(g_{00}\) to the fourth velocity order. From equations~\eqref{eqn:r00ppn} and~\eqref{eqn:r00gauge} we obtain
\begin{eqnarray}
\nabla^2h^{(4)}_{00} &=& -h^{(2)}_{00,i}h^{(2)}_{00,i} + h^{(2)}_{ij}h^{(2)}_{00,ij} - \frac{2\kappa^2}{\Psi_0}\left(v^2 + \frac{\omega_0 + 2}{2\omega_0 + 3}\Pi + \frac{3\omega_0 + 3}{2\omega_0 + 3}\frac{p}{\rho}\right)\rho \nonumber\\
&&+ \frac{m_\psi^2}{2\Psi_0^2 V_2}\left[\left(\frac{\omega_0 + 2}{\Psi_0} + \frac{\omega_1}{2\omega_0 + 3}\right)\rho\psi^{(2)} + (\omega_0 + 2)\rho h^{(2)}_{00}\right]- \frac{\omega_1}{\Psi_0(2\omega_0 + 3)}\psi^{(2)}_{,i}\psi^{(2)}_{,i} \nonumber\\
&&+ \frac{m_\psi^2}{\Psi_0}\left[\left(\frac{2\omega_0 + 1}{2\Psi_0} - \frac{2\omega_1}{2\omega_0 + 3} + \frac{3V_3}{2V_2}\right)\left(\psi^{(2)}\right)^2 -h^{(2)}_{00}\psi^{(2)} + \psi^{(4)}\right] \,.
\end{eqnarray}
Again terms involving \(p\) and \(\Pi\) drop out for the point-like matter source we consider, while terms involving the velocity \(v\) drop out out since we perform the calculation in the rest frame of the matter source. We further neglect the terms \(\rho h_{00}\) and \(\rho\psi\), which correspond to gravitational self-energies and thus contribute only to the potential \(\Phi^{(4)}\) displayed in equation~\eqref{eqn:g00}. The remaining equation can then be integrated and we obtain
\begin{equation}
h^{(4)}_{00}(r) = -2G_{\mathrm{eff}}^2(r)\beta(r)U^2(r)\,,
\end{equation}
where \(\beta(r)\) is given by
\begin{eqnarray}
\beta(r) &=& 1 + \frac{\omega_1e^{-2m_{\psi}r}}{G_{\mathrm{eff}}^2(r)\Psi_0(2\omega_0 + 3)^3} - \frac{m_{\psi}r}{G_{\mathrm{eff}}^2(r)\Psi_0^2(2\omega_0 + 3)}\Bigg[\frac{1}{2}e^{-2m_{\psi}r}\nonumber\\
&&+ \left(m_{\psi}r + e^{m_{\psi}r}\right)\mathrm{Ei}(-2m_{\psi}r) - e^{-m_{\psi}r}\ln(m_{\psi}r)\label{eqn:beta}\\
&&+ \frac{3\Psi_0}{2(2\omega_0 + 3)}\left(\frac{V_3}{V_2} - \frac{1}{\Psi_0} - \frac{\omega_1}{2\omega_0 + 3}\right)\left(e^{m_{\psi}r}\mathrm{Ei}(-3m_{\psi}r) - e^{-m_{\psi}r}\mathrm{Ei}(-m_{\psi}r)\right)\Bigg]\,.\nonumber
\end{eqnarray}
It follows from the asymptotic behaviour of the exponential integral in the case \(x \gg 1\),
\begin{equation}
\mathrm{Ei}(-x) \approx \frac{e^{-x}}{x}\left(1 - \frac{1!}{x} + \frac{2!}{x^2} - \frac{3!}{x^3} + \ldots\right)\,,
\end{equation}
that all terms involving \(\omega_1\) or \(V_3\) fall off proportional to \(e^{-2m_{\psi}r}\), and are thus subleading to the terms involving only \(\omega_0\) and \(V_2\) which fall off proportional to \(e^{-m_{\psi}r}\). We therefore conclude that at large distances \(m_{\psi}r \gg 1\) from the source both \(\omega_1\) and \(V_3\) may be neglected. Again we consider the three limiting cases which we already discussed for \(\gamma\). In the limit \(V_2 \to 0\) and fixed finite \(\omega_0\) we obtain the known result
\begin{equation}
\beta = 1 + \frac{\omega_1\Psi_0}{(2\omega_0 + 3)(2\omega_0 + 4)^2}
\end{equation}
for a massless scalar field~\cite{Nordtvedt1970}. The second case \(1/(2\omega_0 + 3) \to 0\) and fixed finite \(V_2\) yields the limit \(\beta = 1\). We also find the limiting value \(\beta = 1\) in the case \(m_{\psi}r \gg 1\) of a massive scalar field.

This result completes our solution to the post-Newtonian field equations derived in section~\ref{sec:ppn}. We have calculated the metric up to the first post-Newtonian order as displayed in equation~\eqref{eqn:gmunu}. From our calculation we obtained expressions for the effective gravitational constant~\eqref{eqn:geff} and the PPN parameters \(\gamma\)~\eqref{eqn:gamma} and \(\beta\)~\eqref{eqn:beta}. In the next section we will show how these results can be compared to measurements of the PPN parameters, and which restrictions arise on the scalar-tensor theories we consider.


\section{Comparison with experiments}\label{sec:experiments}
In the preceding section we have derived expressions for the PPN parameters \(\gamma(r)\) and \(\beta(r)\) as functions of the expansion coefficients \(\omega_0\), \(\omega_1\), \(V_2\) and \(V_3\). In this section we discuss restrictions on these coefficients which arise from measurements of \(\gamma\) and \(\beta\) in the solar system. We thereby neglect the terms involving \(\omega_1\) and \(V_3\), since their contribution to the PPN parameters at large distances is subleading to that of \(\omega_0\) and \(V_2\), as we argued in the previous section.

The PPN parameters have been measured by various high precision experiments in the solar system~\cite{Will:1993ns,WillReview}. In this article we restrict ourselves to those measurements which provide the strictest bounds on the parameters \(\gamma\) and \(\beta\) and have a characteristic interaction distance \(r_0\). In particular, we will use the bounds obtained from the following experiments:
\begin{itemize}
\item
The deflection of pulsar signals by the Sun has been measured using very long baseline interferometry (VLBI)~\cite{Fomalont:2009zg}. From this \(\gamma\) has been determined to satisfy \(\gamma - 1 = (-2 \pm 3) \cdot 10^{-4}\). Since the radar signals were passing by the Sun at an elongation angle of 3\textdegree, the gravitational interaction distance is \(r_0 \approx 5.23 \cdot 10^{-2}\mathrm{AU}\).
\item
The most precise value for \(\gamma\) has been obtained from the time delay of radar signals sent between Earth and the Cassini spacecraft on its way to Saturn~\cite{Bertotti:2003rm}. The experiment yielded the value \(\gamma - 1 = (2.1 \pm 2.3) \cdot 10^{-5}\). The radio signals were passing by the Sun at a distance of \(1.6\) solar radii or \(r_0 \approx 7.44 \cdot 10^{-3}\mathrm{AU}\).
\item
The most well-known test of the parameter \(\beta\) is the perihelion precession of Mercury~\cite{WillReview}. Its precision is limited by measurements of other contributions to the perihelion precession, most importantly the solar quadrupole moment \(J_2\). The current bound is \(|2\gamma - \beta - 1| < 3 \cdot 10^{-3}\). As the gravitational interaction distance we take the semi-major axis of Mercury, which is \(r_0 \approx 0.387\mathrm{AU}\).
\item
The tightest bounds on \(\beta\) are obtained from lunar laser ranging experiments searching for the Nordtvedt effect, which would cause a different acceleration of the Earth and the Moon in the solar gravitational field~\cite{Hofmann}. For fully conservative theories with no preferred frame effects the current bound is \(4\beta - \gamma - 3 = (0.6 \pm 5.2) \cdot 10^{-4}\). Since the effect is measured using the solar gravitational field, the interaction distance is \(r_0 = 1\mathrm{AU}\).
\end{itemize}
There are more recent experiments which we will not use here since they cannot be characterized by a single value \(r_0\) for the interaction distance. These include in particular combined VLBI measurements of \(\gamma\) at elongation angles between 5\textdegree and 30\textdegree~\cite{Lambert} and measurements of \(\gamma\) and \(\beta\) using ephemeris for a large number of celestial bodies in the solar system~\cite{Fienga:2011qh}. One may argue that also the measurements of \(\gamma\) listed above cannot be characterized by a single interaction distance, since the distance between the Sun and the radio signals varies along their path. A rigorous treatment would therefore require a calculation of null geodesics in the solar system~\cite{Sen}. We will not enter this calculation here and instead assume that the dominant gravitational interaction occurs at the shortest distance to the Sun.

The constraints on the values of the expansion coefficients \(\omega_0\) and \(V_2\) obtained from the experiments listed above can be visualized by plotting the experimentally excluded regions in parameter space, where we use coordinates \(m = 2\kappa\sqrt{\Psi_0V_2}\), measured in inverse astronomical units \(m_{\mathrm{AU}} = 1\mathrm{AU}^{-1}\), and \(\omega_0\). Vertical lines mark regions which are excluded by VLBI measurements, while horizontal lines mark regions which are excluded by Cassini tracking. Similarly, regions with lines from bottom left to top right are excluded by the perihelion precession or Mercury, while regions with lines from top left to bottom right are excluded by lunar laser ranging. Note that the bound from the Cassini experiment is displayed at \(2\sigma\) confidence level, while all other bounds are displayed at \(1\sigma\) confidence level, as we will explain below.

\begin{figure}[hbfp]
\centering
\includegraphics[width=100mm]{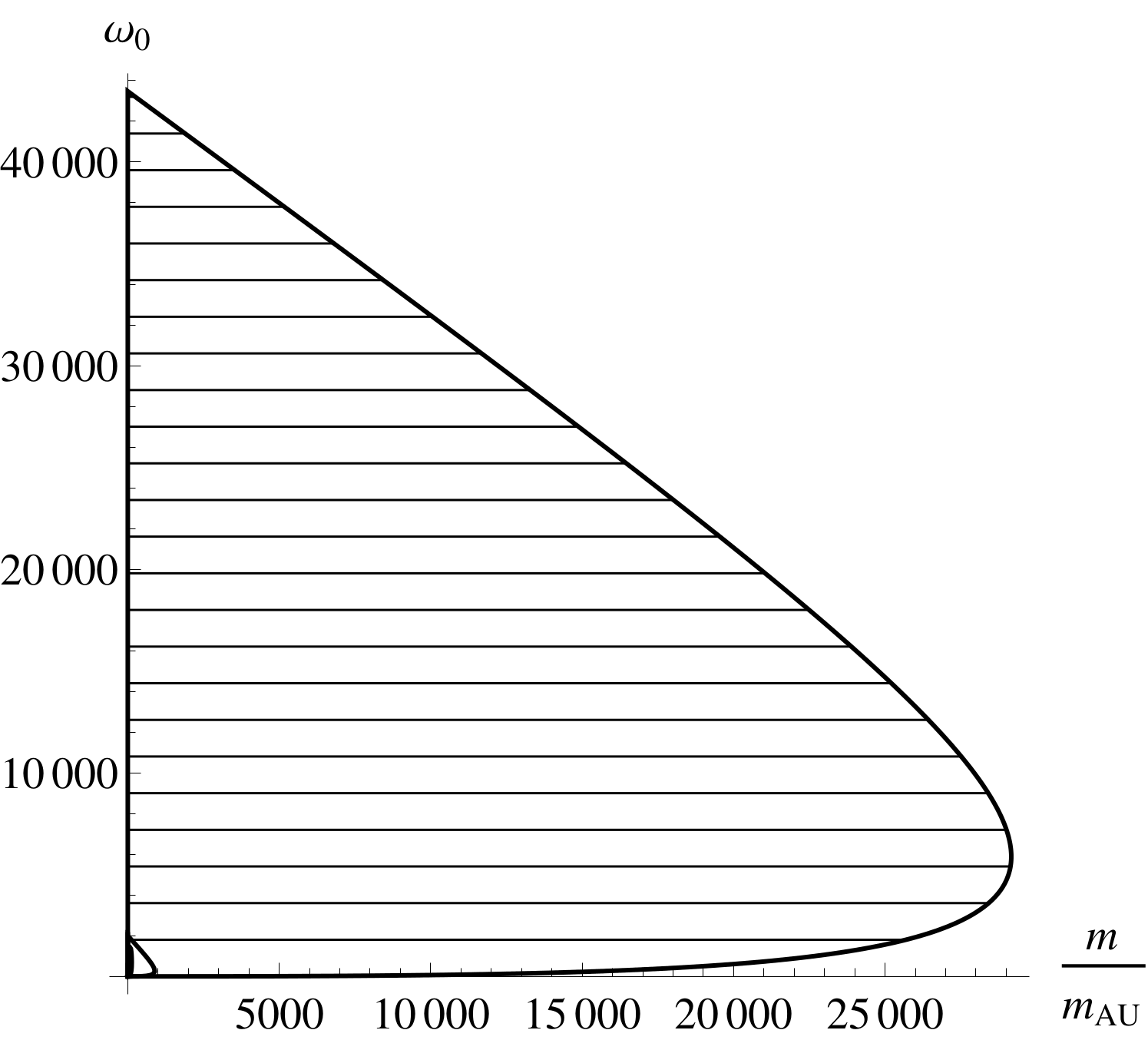}
\caption{Complete excluded region in parameter space.}
\label{fig:full}
\end{figure}

\begin{figure}[hbfp]
\centering
\includegraphics[width=100mm]{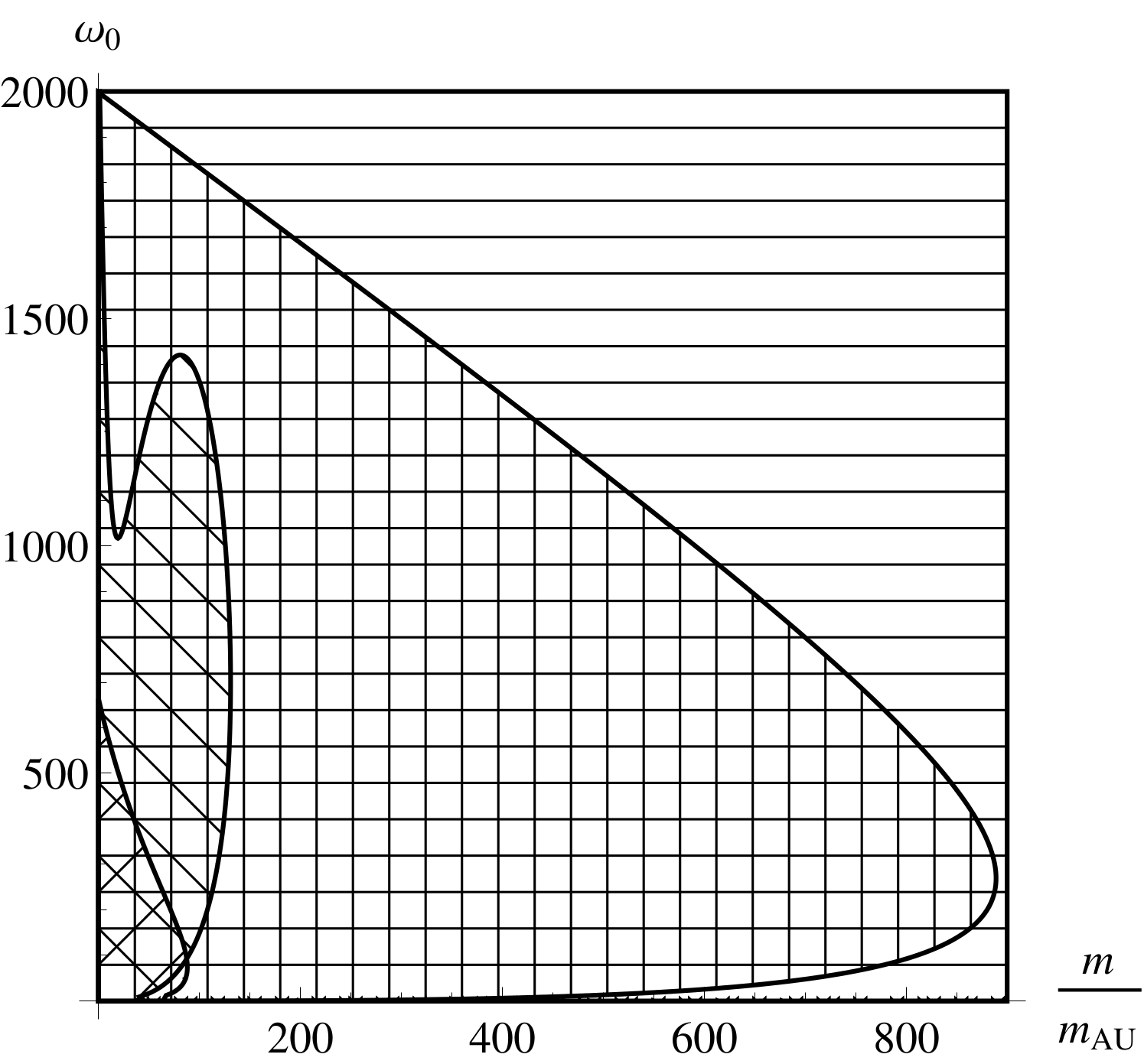}
\caption{Region excluded by VLBI measurements.}
\label{fig:vlbi}
\end{figure}

\begin{figure}[hbfp]
\centering
\includegraphics[width=100mm]{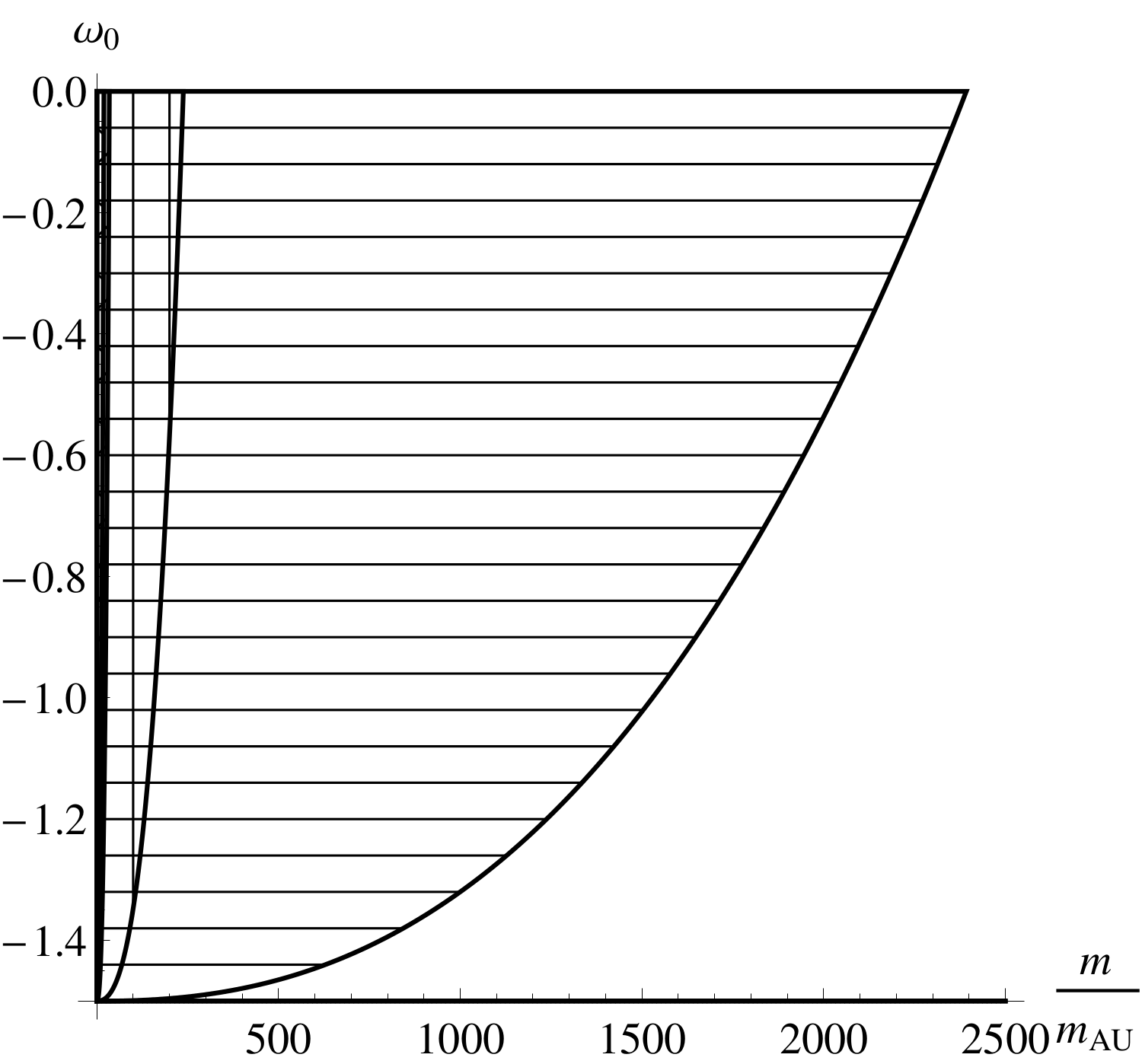}
\caption{Region \(\omega_0 < 0\).}
\label{fig:neg}
\end{figure}

Figure~\ref{fig:full} shows the complete region of the parameter space which is excluded at \(2\sigma\) by the Cassini tracking experiment, from which the tightest bounds on \(\gamma\) have been obtained. As one can see from Fig.~\ref{fig:vlbi} even the \(1\sigma\) bound from VLBI measurements is significantly smaller and fully contained in the region excluded by Cassini. This can also be seen for negative values of \(\omega\) from Fig.~\ref{fig:neg}.

\begin{figure}[hbfp]
\centering
\includegraphics[width=100mm]{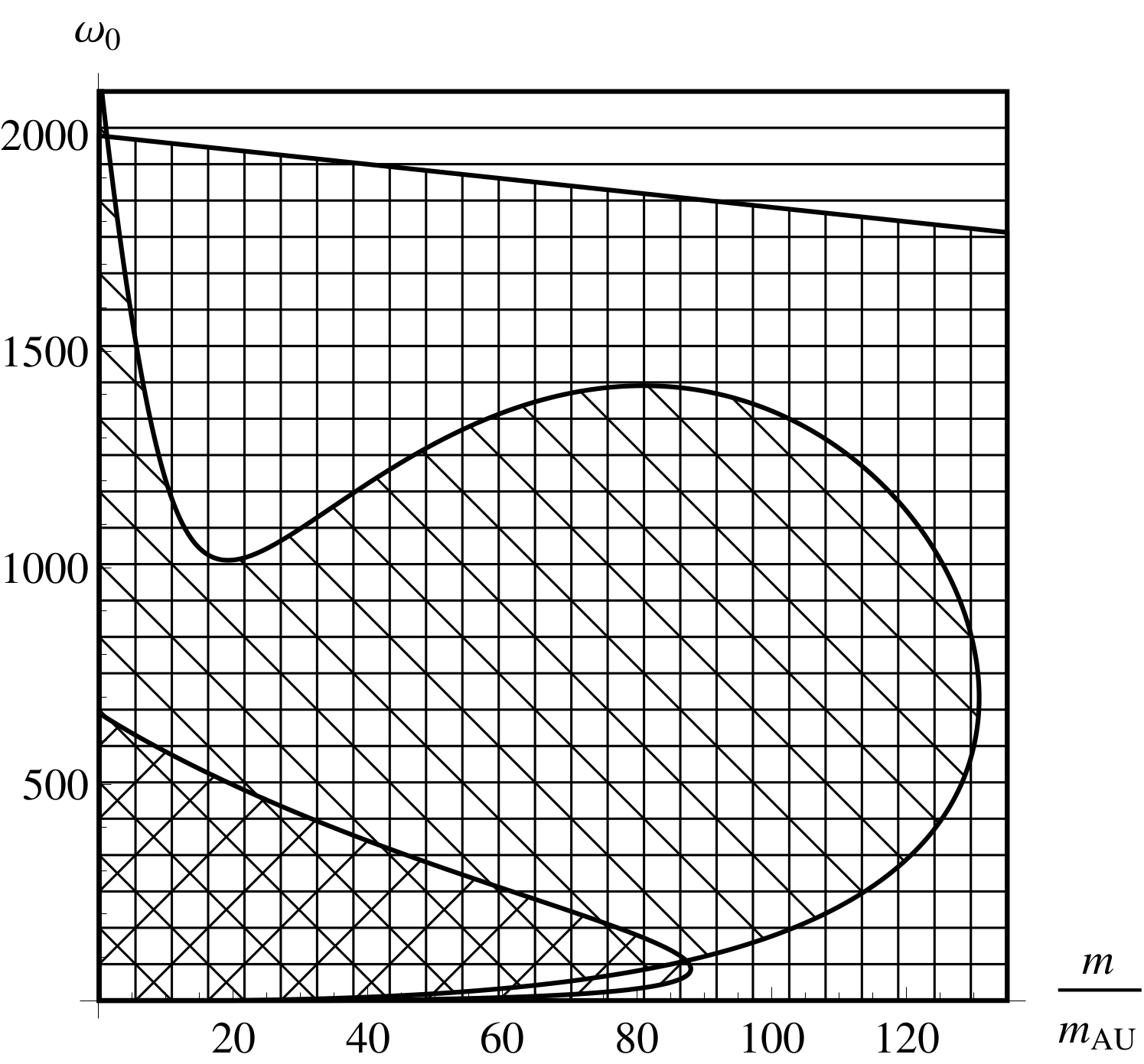}
\caption{Constraints from \(\beta\).}
\label{fig:beta}
\end{figure}

\begin{figure}[hbfp]
\centering
\includegraphics[width=100mm]{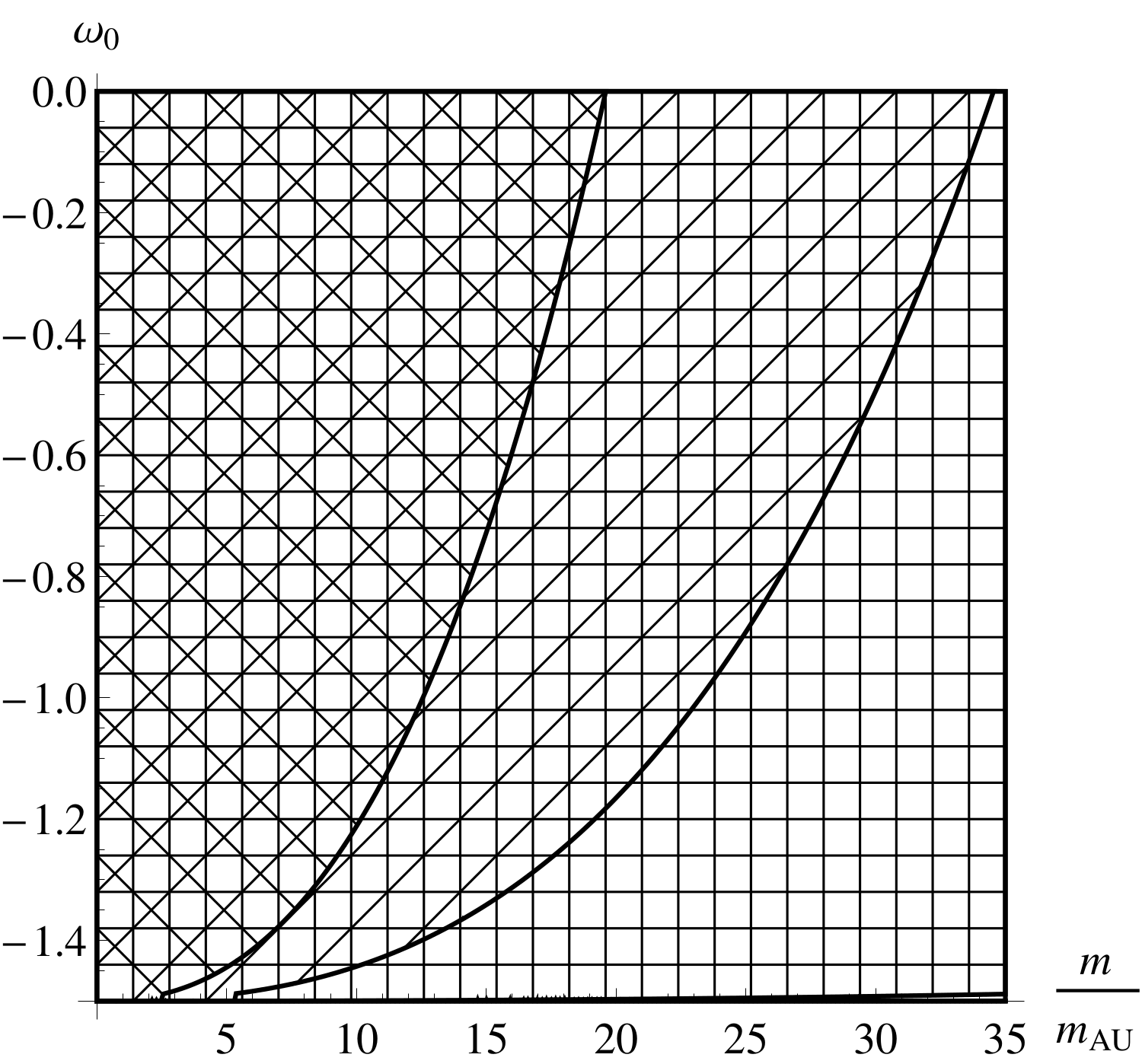}
\caption{Constraints from \(\beta\) for \(\omega_0 < 0\).}
\label{fig:betaneg}
\end{figure}

The \(1\sigma\) bounds obtained from measurements of \(\beta\) are shown in Fig.~\ref{fig:beta}. One can see that they are located entirely in the region already excluded by the Cassini measurement of \(\gamma\). This also holds for negative values of \(\omega_0\) as shown in Fig.~\ref{fig:betaneg}. This result is due to the fact that the bounds on \(\beta\) are significantly weaker than the bounds on \(\gamma\). In order to obtain stricter constraints on the parameter space from \(\beta\), experiments at higher precision are required. Alternatively measurements of \(\gamma\) or \(\beta\) at smaller interaction distance \(r_0\) could exclude regions of the parameter space for larger values of $m$.

\section{Conclusion}\label{sec:discussion}
In this article we discussed the post-Newtonian approximation of scalar-tensor theories of gravity with arbitrary coupling function $\omega(\Psi)$ and arbitrary potential $V(\Psi)$. In the post-Newtonian limit these functions can be characterized by four coefficients \(\omega_0, \omega_1, V_2, V_3\) of the Taylor expansion. We then defined the PPN parameters \(\gamma\) and \(\beta\) and calculated their values for a static point mass as functions of these four constants. It turned out that the PPN parameters are not constant, but depend on the distance between the gravitating source and the test mass. In the appropriate limits our expressions reduce to earlier results obtained for simpler cases \cite{Nordtvedt1970, Olmo:2005hc, Perivolaropoulos:2009ak}. We further found that for a massive scalar field the contribution of \(\omega_1\) and \(V_3\) to the values of the PPN parameters can be neglected at large distances from the source, $m_\psi r\gg 1$.

We finally compared our results to measurements of \(\gamma\) and \(\beta\) in the solar system and obtained bounds on the constants \(\omega_0\) and \(V_2\). It turned out that the strictest bounds are obtained from measurements of \(\gamma\) by Cassini tracking, so that the current measurements of \(\beta\) place no further restrictions on \(\omega_0\) and \(V_2\). In order to obtain new bounds it would be necessary to measure the parameters \(\gamma\) and \(\beta\) at higher precision or at shorter interaction distances between the gravitational source and the test mass. The latter would allow probing and possibly excluding larger values of \(V_2\), corresponding to a higher mass of the scalar field.

The work presented in this article allows for various further studies. While we have restricted ourselves to a calculation of the PPN parameters \(\gamma\) and \(\beta\) for a static point mass, one may ask which further contributions to the post-Newtonian metric arise from moving sources or sources with non-vanishing pressure, internal energy or gravitational self-energy. A calculation of these contributions would provide insight into further PPN parameters, which are related to preferred-frame and preferred-location effects. Comparison of these parameters with their values measured in the solar system may place further bounds on viable theories of scalar-tensor gravity. Another line of investigation would be the case of strong fields provided by, e.g., binary pulsars. We further aim to extend our work to theories with more than one scalar degree of freedom, and thus chart the landscape of multi-scalar-tensor gravity.

\acknowledgments
The authors would like to thank Andreas Sch\"arer for pointing out an error in a previous version of this manuscript.
This work was supported by the Estonian Research Council
Institutional Research Funding Grant No. IUT02-27,
by the Estonian Science Foundation Grant No. 8837
and by the European Union through the European Regional
Development Fund (Centre of Excellence TK114). MH gratefully acknowledges full financial support
from the Estonian Research Council through the
Postdoctoral Research Grant ERMOS115.

\end{document}